\documentclass[compsoc,conference,a4paper,10pt,times]{IEEEtran}
\IEEEoverridecommandlockouts
\usepackage{cite}
\usepackage{amsmath,amssymb,amsfonts}
\usepackage{algorithmic}
\usepackage{graphicx}
\usepackage{textcomp}
\usepackage{bmpsize}
\usepackage{wasysym}
\usepackage{pifont}
\usepackage{xcolor}
\usepackage{lipsum}
\usepackage[colorlinks=true,urlcolor=black]{hyperref}
\usepackage{booktabs}
\usepackage{tabularx}
\usepackage{enumitem} 
\usepackage{xcolor}
\usepackage{tabularx}
\usepackage{booktabs}
\usepackage{array}
\usepackage{caption}
\usepackage{makecell}

\usepackage{booktabs}
\usepackage{array}
\usepackage[table,xcdraw]{xcolor}

\usepackage[most]{tcolorbox}

\newcommand{\cmark}{\ding{51}}   
\newcommand{\starx}{\ding{72}}   
\newcommand{\pmark}{\LEFTCIRCLE} 
\def\BibTeX{{\rm B\kern-.05em{\sc i\kern-.025em b}\kern-.08em
    T\kern-.1667em\lower.7ex\hbox{E}\kern-.125emX}}
\begin{document}

\title{SoK: Privacy-aware LLM in Healthcare: Threat Model, Privacy Techniques, Challenges and Recommendations\\ 
}

\author{
\IEEEauthorblockN{
Mohoshin Ara Tahera\IEEEauthorrefmark{1},
Karamveer Singh Sidhu\IEEEauthorrefmark{2},
Shuvalaxmi Dass\IEEEauthorrefmark{1},
Sajal Saha\IEEEauthorrefmark{2}
}
\IEEEauthorblockA{\IEEEauthorrefmark{1}University of Louisiana at Lafayette, Lafayette, LA, USA}
\IEEEauthorblockA{\IEEEauthorrefmark{2}University of Northern British Columbia, Canada}

\IEEEauthorblockA{
Emails: \texttt{mohoshin-ara.tahera1,shuvalaxmi.dass@louisiana.edu}\\
\texttt{ksidhu,sajal.saha@unbc.ca}
}
}


\maketitle

\begin{abstract}
Large Language Models (LLMs) are increasingly adopted in healthcare to support clinical decision-making, summarize electronic health records (EHRs), and enhance patient care. However, this integration introduces significant privacy and security challenges, driven by the sensitivity of clinical data and the high-stakes nature of medical workflows. These risks become even more pronounced across heterogeneous deployment environments, ranging from small on-premise hospital systems to regional health networks, each with unique resource limitations and regulatory demands.
This Systematization of Knowledge (SoK) examines the evolving threat landscape across the three core LLM phases: Data preprocessing, Fine-tuning, and Inference within realistic healthcare settings. We present a detailed threat model that characterizes adversaries, capabilities, and attack surfaces at each phase, and we systematize how existing privacy-preserving techniques (PPTs) attempt to mitigate these vulnerabilities.
While existing defenses show promise, our analysis identifies persistent limitations in securing sensitive clinical data across diverse operational tiers. We conclude with phase-aware recommendations and future research directions aimed at strengthening privacy guarantees for LLMs in regulated environments. This work provides a foundation for understanding the intersection of LLMs, threats, and privacy in healthcare, offering a roadmap toward more robust and clinically trustworthy AI systems.

\end{abstract}


\section{Introduction}
LLMs are increasingly integrated into healthcare for clinical documentation, decision support, radiology and pathology report summarization, and clinician–patient communication (representative applications in Table \ref{tab:healthcare-application}). Deployments span radiology/report summarization \cite{paper020,reference84}, triage and clinical decision support (CDS) systems \cite{reference10,paper016}, and multilingual dialogue assistants \cite{paper04,reference14}. While these applications demonstrate substantial utility, they also introduce significant privacy risks due to the sensitivity of Protected Health Information (PHI) and stringent regulatory constraints.

This SoK examines privacy concerns across the three core operational phases of LLMs: data preprocessing, fine-tuning, and inference, with a specific focus on text-centric healthcare applications (e.g., EHR notes, discharge summaries, DICOM-derived reports, pathology and radiology narratives, and clinician–patient transcripts). We intentionally scope the SoK to text-generating or text-consuming LLM pipelines, excluding imaging-only models unless they interact with PHI-bearing textual artifacts.

Unlike prior surveys \cite{2023privacy,reference64,2025dataprivacy,paper06,aljohani2025comprehensive,2025medllm}, which primarily catalog techniques or discuss privacy at a high level, our work provides a phase-aligned systematization. We unify terminology and explicitly map:
attack surfaces→enabled attacks→defenses→remaining limitations
grounded in healthcare data artifacts such as EHR/HL7/FHIR structures, DICOM text fields, and clinical transcripts. Prior work does not systematically link adversary capabilities to specific vulnerabilities across the three phases; Table \ref{tab:survey-comparison-healthcare} highlights these gaps. Our corpus spans peer-reviewed studies and authoritative preprints from 2020–2025 involving clinical datasets, hospital deployments, or consortium-based learning; details appear in section \ref{sec:Literature_Collection}.

This SoK answers three guiding research questions in each phase:
\begin{itemize}
    \item \textbf{RQ1.}How should adversaries in healthcare LLMs be categorized (e.g., internal vs. external), and how do their capabilities and prior knowledge shape the attack vectors that emerge at each phase of the LLM lifecycle?
    \item \textbf{RQ2.} How effectively do current privacy-preserving techniques mitigate phase-specific vulnerabilities?
    \item \textbf{RQ3.} What limitations remain in current defenses, and what phase-aware strategies can guide future privacy-enhanced LLM development?
\end{itemize}
To address \textbf{RQ1}, we construct a detailed threat model for each phase, identifying adversary capabilities, prior knowledge, and corresponding vulnerabilities. For \textbf{RQ2}, we evaluate privacy-preserving techniques and analyze their effectiveness relative to the identified phase-specific threats. For \textbf{RQ3}, we synthesize limitations and propose future directions and recommendations for phase-aware, threat-resilient, healthcare-specific privacy enhancements.

Given that healthcare deployments commonly rely on distributed architectures, our analysis of the fine-tuning stage adopts a Federated Learning (FL) framework, emphasizing vulnerabilities and defenses in the Federated Fine-Tuning Phase.
By answering these questions, we make the following three contributions:

\begin{enumerate}
\item \textbf{Phase-Specific Threat Model:}
We develop a comprehensive threat model for healthcare LLMs, categorizing adversaries by location (internal vs. external) and capability and identifying attack surfaces and key vulnerabilities across data preprocessing, federated fine-tuning, and inference (e.g., gradient leakage, internal update exposure, model extraction).
\item \textbf{Evaluation of Privacy-Preserving Techniques:}
We systematically analyze defenses such as differential privacy, secure aggregation, inference-time mitigations and evaluate how effectively they address the vulnerabilities surfaced in each phase. Our findings show that existing techniques provide partial protection but are often phase-agnostic or misaligned with healthcare-specific threats.

\item \textbf{Limitations and Future Directions:}
We identify critical limitations in current privacy strategies and propose future research directions aimed at developing phase-aware, threat-resilient, and privacy-enhanced LLMs tailored for healthcare environments. These include recommendations such as standardized data anonymization, adaptive differential privacy to mitigate gradient leakage while preserving rare-disease fidelity in federated fine-tuning.
\end{enumerate}

Following the literature collection methodology (Section \ref{sec:Literature_Collection}), our SoK is structured into three sections, each focused on a distinct phase of the LLM lifecycle: Data Preprocessing (Section \ref{sec:data_prepocessing}), Federated Fine-Tuning (Section \ref{sec:fine_tuning}), and Inference (Section \ref{sec:inference}). Each section includes two subsections: The Threat Model subsection addresses \textbf{RQ1}, outlining adversaries, capabilities, and attack vectors. The Privacy Preserving Defenses subsection covers the  rest of the RQs: \textbf{RQ2} with \textit{Takeaways} on how existing privacy-preserving techniques mitigate vulnerabilities, and \textbf{RQ3} with \textit{Limitations} of current defenses and \textit{Recommendations}. 
To synthesize insights across phases, we present Tables~\ref{tab:tab1},~\ref{tab:tab2} and ~\ref{tab:tab3}, which systematically maps the threats and vulnerabilities of each phase to their corresponding defenses, limitations, and recommendations, offering a concise, phase-specific roadmap for privacy-aware LLMs.

\begin{table*}[ht]
\centering
\caption{Representative Healthcare Applications of LLMs with example tasks}
\label{tab:healthcare-application}

\renewcommand{\arraystretch}{1.5} 

\begin{tabular}{p{0.32\linewidth} p{0.62\linewidth}}
\toprule
\textbf{Application Type} & \textbf{Example Tasks} \\
\midrule
\includegraphics[width=0.08\linewidth]{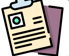} \, Electronic Health Records (EHR) 
& Demographics, diagnoses, labs, ICU notes \& pathology (MIMIC, n2c2) \cite{reference01,reference02,reference12}; insurance claims \cite{reference10}; rare cohorts \cite{paper0034}; synthetic EHRs for comorbidity/survival modeling \cite{reference26,reference36,reference37} \\

\includegraphics[width=0.08\linewidth]{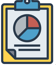} \, Clinical Summaries \& Documentation 
& Discharge-note assistants \cite{reference16,paper016}; cardiac/oncology summaries \cite{paper0034,paper03}; multilingual notes \cite{paper04,paper011}; ICU timelines \cite{reference57}; leakage risks (telemedicine) \cite{paper033,paper034} \\

\includegraphics[width=0.08\linewidth]{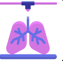} \, Medical Imaging \& Signal Data 
& Radiology summarization (CT, X-ray) \cite{paper020,reference84}; pathology summarizers \cite{paper03,reference14}; ChestX-ray14 \cite{reference64}; cardiology data (ECG, BP, cholesterol, EF) \cite{reference16,reference84} \\

\includegraphics[width=0.08\linewidth]{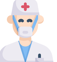} \, Clinical Decision Support (CDS) 
& Rule-based CDS (guidelines) \cite{reference35}; oncology classification \cite{paper03,reference15}; cardiovascular outcome prediction \cite{reference10,paper016}; comorbidity prediction (graph prompting, RAG) \cite{reference37}; multimodal support \cite{paper05,reference15} \\

\includegraphics[width=0.08\linewidth]{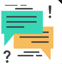} \, Telemedicine \& Patient Interaction 
& Triage chatbots \cite{paper028,paper026}; telemedicine dialogues \cite{paper033}; multilingual transcripts \cite{paper04,reference14}; DSS for rare tumor queries \cite{paper019,paper022}; mobile health apps (chest pain triage) \cite{paper025,reference16} \\
\bottomrule
\end{tabular}
\end{table*}

\begin{table*}[t]
\centering
\caption{Comparison of healthcare-focused surveys vs. our SoK. 
Legend: \starx = strong/unique/comprehensive; \cmark = present/addressed; $\pmark$ = Partial; $\ocircle$ = Narrow/absent; Full = Pre-Processing → Federated Fine-Tuning → Inference}
\label{tab:survey-comparison-healthcare}

\renewcommand{\arraystretch}{1.8} 
\footnotesize 
\setlength{\tabcolsep}{5pt} 

\begin{tabular}{p{2.8cm} p{1.3cm} p{1.3cm} p{1.3cm} p{1.3cm} p{1.3cm} p{1.3cm} p{3.2cm}}
\toprule
\textbf{Criteria} & \textbf{\cite{2023privacy}} & \textbf{\cite{reference64}} & \textbf{\cite{2025dataprivacy}} & \textbf{\cite{paper06}} & \textbf{\cite{aljohani2025comprehensive}} & \textbf{\cite{2025medllm}} & \textbf{Our SoK} \\
\midrule
\textbf{Stages Covered} & \pmark Partial & \pmark Partial & \pmark Partial & \pmark Partial & \pmark Partial & \pmark Partial & \starx Full \\
\textbf{Threat Model} & $\ocircle$ Narrow & $\ocircle$ Narrow & $\ocircle$ Narrow & $\ocircle$ Narrow & $\pmark$ Partial & $\pmark$ Partial & \starx Comprehensive, phase-mapped \\
\textbf{Defenses / PPTs} & \cmark Partial & \cmark Partial & $\ocircle$ Narrow & \cmark Partial & \cmark Partial & \cmark Partial & \starx Tailored + Phase-mapped \\
\textbf{Limits of Defenses} & \cmark Present & $\ocircle$ Absent & $\ocircle$ Absent & $\ocircle$ Absent & \cmark Present & \cmark Present & \starx Phase-tied \\
\textbf{Recommendations} & \cmark Present & \cmark Present & \cmark Present & $\ocircle$ Absent & \cmark Present & \cmark Present & \starx Gap-tied, phase-specific \\
\bottomrule
\end{tabular}
\normalsize
\end{table*}

\section{Literature Collection}
\label{sec:Literature_Collection}
This SoK follows a structured but targeted methodology aimed at systematizing privacy and security risks for healthcare LLMs across the three operational phases: data preprocessing, federated fine-tuning, and inference, prioritizing works that directly engage with privacy, security, or compliance rather than providing exhaustive coverage of all clinical LLM applications. We searched IEEE Xplore, ACM DL, SpringerLink, ScienceDirect, PubMed, and major preprint repositories (arXiv, medRxiv) for the years 2019–2025 using combined terms spanning model types \textit{(“LLM,” “large language model,” “foundation model”)}, healthcare domains \textit{(“medical,” “clinical,” “EHR,” “radiology,” “pathology,” “telemedicine”)}, and privacy/security mechanisms \textit{(“privacy,” “differential privacy,” “federated learning,” “secure aggregation,” “homomorphic encryption,” “access control”)}. Searches were supplemented with backward and forward snowballing from influential surveys on LLM privacy and healthcare AI. We included papers that (i) involve LLMs or closely related foundation models or core privacy-preserving mechanisms relevant to healthcare (e.g., DP, FL, HE/MPC, anonymization, access control); (ii) use or explicitly target healthcare data or clinical workflows such as EHRs, imaging, pathology, telemedicine, or medical QA; and (iii) provide a substantive treatment of privacy, security, or regulatory constraints, including threat models, attacks, defenses, or evaluations. We excluded purely application-focused clinical LLM papers with only passing references to privacy, as well as generic privacy/security works lacking a clear connection to healthcare or LLMs.

\section{Data Preprocessing Phase}
\label{sec:data_prepocessing}
The preprocessing phase involves cleaning, structuring, and transforming raw clinical data to address issues like missing values, feature normalization, class imbalance, and skewed distributions \cite{reference01, reference02}. In healthcare, this phase is critical due to the sensitivity and diversity of data from EHRs, medical imaging, claims, and patient-generated sources.

\subsection{Threat Model}
The data preprocessing stage ingests heterogeneous clinical data including EHR tables, radiology metadata, lab results, claims records, and patient-generated content—where identifiers and quasi-identifiers remain intact \cite{reference05, breference1}. Because adversaries interact with the raw substrate from which LLM datasets are formed, this phase presents uniquely powerful opportunities for privacy compromise and data manipulation \cite{reference49, reference79}.

\subsubsection{Attacker Landscape and Incentives}
Preprocessing threats stem from both internal and external adversaries, whose incentives and privileges shape distinct privacy risks.

\textbf{Internal adversaries} are the most operationally impactful, as they handle raw data throughout daily workflows.
\textbf{(1)} Clinical staff (nurses, residents, and attending physicians) may export EHR data for shift handovers or audits through informal channels, unintentionally exposing unmasked identifiers and clinical notes \cite{breference1}.
\textbf{(2)} Data engineers and ML researchers maintaining ETL and data-cleaning pipelines often retain identifiers for debugging or convenience, propagating PHI leakage into intermediate tables and logs \cite{breference3,etl_security_healthcare2024,reference94,reference95}.
\textbf{(3)} System administrators and IT personnel possess privileged access to storage servers, backups, and preprocessing nodes. Misconfigured access controls or network shares can silently expose PHI to unauthorized internal users or external exploits \cite{etl_security_healthcare2024}.

\textbf{External adversaries} target the same infrastructures but with financial, strategic, or competitive motives.
\textbf{(1)} Ransomware groups exploit unpatched or misconfigured ETL servers and data lakes, as illustrated by the WannaCry attack on the NHS \cite{wannaCryNHS}.
\textbf{(2)} Corporate or state-linked attackers infiltrate healthcare data lakes to harvest PHI for analytics or commercial leverage \cite{reference49,2025dataprivacy}.
\textbf{(3)} Research competitors may compromise preprocessing APIs or storage connectors to reconstruct cohorts or infer institutional practices \cite{paper015,reference67,breference16}.

\subsubsection{Prior Knowledge and Capability Gradient}
Adversaries in this phase often hold strong prior knowledge of healthcare data systems, making even limited access highly dangerous, such as:
\textbf{(1)} They understand EHR schemas and identifiers (HL7/FHIR fields, MRNs, timestamps, and ICD codes), enabling precise targeting of PHI within raw exports \cite{reference01,reference02,breference1}.
\textbf{(2)} They exploit quasi-identifiers such as age, location, ethnicity, or rare conditions for re-identification and linkage before anonymization \cite{reference05,narayanan2008robust}.
\textbf{(3)} They are familiar with ETL workflows and data handling practices, including where staging files, failed job outputs, and temporary exports are stored \cite{breference3,etl_security_healthcare2024}.
\textbf{(4)} Such insider knowledge directly enables exploitation of weakly protected APIs and misconfigured file systems that connect preprocessing tools to hospital databases.
\textbf{(5)} High-resource externals also combine auxiliary datasets such as insurance claims, leaked registries, and prior hospital breaches to strengthen re-identification or model poisoning efforts \cite{stadler2022synthetic}.

The capability spectrum spans: 
\textbf{(1)} low-resource insiders (e.g., nurses, clerks) can unintentionally leak identifiers through manual exports or shift handover lists. \cite{breference1,reference05};
\textbf{(2)} moderate-resource insiders (e.g., ETL engineers, ML staff) handle staging data and debugging logs containing unmasked PHI. \cite{reference94,reference95}; and 
\textbf{(3)} high-resource externals (e.g., ransomware groups, APTs) exploit insecure APIs, exfiltrate backups, or inject poisoned HL7/FHIR records to compromise downstream model integrity. \cite{a7,jagielski2018poisoning}.

\subsubsection{Attack Surface Vulnerabilities and Enabled Attacks}
These layered attacker capabilities map directly to the core vulnerabilities of the preprocessing stage:
\begin{itemize}[noitemsep, topsep=0pt, leftmargin=*]
    \item \textbf{Quasi-identifiers:} Demographic and clinical combinations exploited by insiders and externals for re-identification \cite{reference05}.
    \item \textbf{Data leakage via logs:} Debug outputs and ETL failure logs containing PHI accessible to authorized insiders or compromised systems \cite{reference95}.
    \item \textbf{Weak access controls and APIs:} Misconfigured permissions, exposed endpoints, or insecure ETL connectors exploited by administrators or external attackers to access raw datasets \cite{breference3,etl_security_healthcare2024}.
\end{itemize}

These vulnerabilities directly enable several concrete attacks: \textbf{(1) Re-identification/linkage attacks}, where adversaries combine quasi-identifiers with auxiliary records to recover patient identities \cite{narayanan2008robust}. \textbf{(2) Log-based PHI leakage}, where unmasked identifiers in ETL traces or debug outputs expose raw clinical details \cite{reference94}. \textbf{(3) Unauthorized dataset access}, where weak permissions or misconfigured APIs allow direct retrieval of raw EHR exports or staging files \cite{etl_security_healthcare2024}. \textbf{(4) Data poisoning}, where attackers inject manipulated or fabricated clinical records into HL7/FHIR streams to distort downstream fine-tuning behavior \cite{jagielski2018poisoning}. 

Together, these attacks operationalize the incentives, capabilities, and prior knowledge described in the preprocessing threat model, demonstrating how seemingly routine workflow weaknesses become attack vectors.

\begin{table*}[t]
\centering
\caption{Data Preprocessing Phase — Summary of Threats, Defenses, Limitations, and Recommendations.}
\label{tab:preprocessing-taxonomy}

\renewcommand{\arraystretch}{1.25}
\footnotesize
\setlength{\tabcolsep}{3.8pt}

\begin{tabularx}{\textwidth}{|>{\raggedright\arraybackslash}p{4cm}|
                                >{\raggedright\arraybackslash}p{4cm}|
                                >{\raggedright\arraybackslash}p{4cm}|
                                >{\raggedright\arraybackslash}p{4cm}|}
\hline
\rowcolor[HTML]{E8E8E8}
\textbf{Threat Model} &
\textbf{Defenses (PPTs)} &
\textbf{Limitations} &
\textbf{Recommendations} \\
\hline

\begin{itemize}[leftmargin=*,nosep]
    \item \textbf{Adversaries:} Internal (clinicians, data engineers, IT admins) and external (ransomware, state-linked, competitors).
    \item \textbf{Capabilities:} Exploit quasi-identifiers, extract PHI from ETL logs/debug traces, abuse weak APIs, or inject poisoned HL7/FHIR data.
    \item \textbf{Key Vulnerabilities:} Quasi-identifiers, PHI-bearing logs, weak access control, raw exports/backups.
\end{itemize}
&
\begin{itemize}[leftmargin=*,nosep]
    \item \textbf{Anonymization:} Masking, tokenization, k-anonymity, l-diversity, pseudonym vaults.
    \item \textbf{Synthetic Data:} GAN/VAE/LLM-based generation replacing real EHR inputs.
    \item \textbf{Differential Privacy:} DP-SGD, local DP, stochastic embedding noise for pre-training or tabular features.
\end{itemize}
&
\begin{itemize}[leftmargin=*,nosep]
    \item \textbf{Quasi-identifiers:} Generators may reproduce unique attribute patterns, enabling linkage.
    \item \textbf{Logs / APIs:} PHI may persist in debug traces; DP misconfiguration leaks identifiers.
    \item \textbf{Poisoning:} Defenses fail against manipulated records injected pre-noise or masking.
    \item \textbf{Governance:} Weak auditing or inconsistent DP parameters reintroduce vulnerabilities.
\end{itemize}
&
\begin{itemize}[leftmargin=*,nosep]
    \item \textbf{Quasi-identifiers:} Structured anonymization within ETL (tokenization, suppression).
    \item \textbf{Log leakage:} LLM-assisted scrubbing of debug/staging data.
    \item \textbf{Access control:} Least-privilege, vault-based tokens and synthetic sandboxes.
    \item \textbf{Poisoning:} Pre-ingestion anomaly detection and HL7/FHIR integrity checks.
\end{itemize}
\\
\hline
\end{tabularx}
\label{tab:tab1}
\end{table*}

\subsection{Privacy-Preserving Defenses}
Given the risks of handling PHI, implementing robust privacy-preserving techniques in this phase is crucial \cite{reference03}. Common methods like \textit{data anonymization}, \textit{synthetic data generation}, and \textit{noise addition} help safeguard patient privacy while enabling LLM training. 

\subsubsection{Data Anonymization} 
It is the first operational defense in preprocessing, directly countering insider misuse and quasi-identifiers exposure before clinical data enter modeling pipelines. During cleaning, raw EHR tables, radiology metadata, and lab files still carry names, MRNs, timestamps, and geocodes, which are the prime targets for careless or malicious insiders \cite{breference1, breference3, reference94, reference95}. As outlined in the threat model, low-resource insiders (nurses, clerks) leak unmasked exports in handovers, and moderate-resource insiders (ETL/ML staff) surface identifiers in debug logs and staging data \cite{etl_security_healthcare2024}. Early masking or tokenization interrupts these leakage paths so exports, logs, and cached views hold only pseudonymous values even when systems are misconfigured.

Classical frameworks such as k-anonymity, l-diversity, t-closeness generalize or perturb demographic/clinical fields so attribute combinations cannot uniquely identify a patient \cite{reference05, reference06, reference07}. This directly mitigates the linkage risks noted in the threat model, where adversaries combine auxiliary attributes (age, ZIP, rare conditions) to re-identify patients \cite{reference05, narayanan2008robust}. Hospitals commonly truncate postal codes or merge rare diagnoses (e.g., oncology registries), blunting the advantage of external actors who leverage demographic priors or leaked registries \cite{stadler2022synthetic}. Complementary pseudonymization and encryption replace identifiers with reversible tokens kept in audited vaults \cite{reference08, reference09}, constraining the practical privileges of ETL personnel and admins: if staging/backup files are accessed, only encrypted tokens, not PHI, are exposed, cutting off the same audit log and backup channels exploited by insider error or lateral movement \cite{reference94, reference95}.

\textbf{Using anonymized datasets.} 
Most healthcare-LLM studies lean on pre-anonymized corpora (MIMIC-III/IV, MIMIC-CXR) de-identified for HIPAA/GDPR \cite{reference01, reference02, reference04, reference08}, enabling clinical IE \cite{reference10}, privacy-preserving local deployment \cite{reference11}, and federated pretraining \cite{reference12}. Yet dataset-level compliance does not eliminate pipeline-level exposure: identifiers can resurface in temporary exports, logs, or monitoring dashboards during hospital preprocessing \cite{reference94}, and quasi-identifiers can still support attribute inference or record linkage \cite{fredrikson2015model, paper015}. Some studies instead use closed-source anonymized sets \cite{reference13} or “PII-free by design” resources (ChatDoctor, MMQS, IMCS-21) \cite{reference14, reference18}, which avoid direct identifiers but rarely provide end-to-end audit evidence that PHI never entered the pipeline.

\textbf{Using LLMs for anonymization.} 
An emerging mitigation embeds anonymization inside preprocessing itself: LLMs fine-tuned for medical de-identification redact PHI while preserving clinical semantics \cite{reference21}. These tools address the free-text/logging vulnerabilities highlighted in the threat model, e.g., identifiers buried in notes or timestamps that leak via ETL logs and debug outputs \cite{breference3, reference94} and, when paired with hybrid NER–DL pipelines, sanitize text streams before storage or model ingestion, closing insider and scraper pathways that rely on residual identifiers for linkage \cite{stadler2022synthetic}. Implementation quality, however, is uneven: some datasets report manual redaction (e.g., HajjHealthQA) \cite{reference15}, others assume PHI absence in synthetic dialogues \cite{reference23, reference24}, and several open resources still show explicit identifiers without automated scrubbing \cite{reference25}. This inconsistency reinforces the need to treat anonymization as an engineered control in the preprocessing pipeline, not merely a property of upstream datasets.

\textcolor{blue}{\textbf{Takeaway.}}
Anonymization directly mitigates the preprocessing threats by disrupting how insiders and externals exploit raw clinical data before transformation.
Masking or encrypting identifiers before ETL prevents \textit{re-identification} and \textit{linkage attacks} via quasi-identifiers such as age or rare conditions, while also blocking \textit{log-based PHI leakage} from debug outputs \cite{reference05,reference94}.
For \textit{high-resource external adversaries}, it removes identifiers that enable cross-record matching through their prior knowledge and auxiliary datasets, thereby reducing the utility of stolen clinical data. Consistent application of this technique substantially reduces exposure to core preprocessing vulnerabilities: \textit{quasi-identifiers misuse, log leakage, and weak access controls}, forming a crucial first layer of defense for privacy-preserving LLM pipelines.

\textcolor{violet}{\textbf{Limitation.}
While anonymization mitigates key preprocessing threats, it leaves several vulnerabilities partially exposed. \textit{(1) Quasi-identifiers:}it cannot fully prevent re-identification or linkage attacks by high-resource externals leveraging prior knowledge and auxiliary datasets; \textit{(2) Data leakage via logs:} when applied only at the dataset level, it fails to mitigate PHI exposure through debug outputs, temporary exports, or audit traces, allowing low- and moderate-resource insiders to access identifiers left in system logs or misconfigured dashboards. \textit{(3) Weak access controls:} anonymization does not prevent data poisoning or unauthorized input manipulation, as moderate-resource insiders can still inject or alter records before masking through insecure ETL access points.}

\subsubsection{Synthetic Data Generation} 
This operates as a proactive privacy defense in the preprocessing phase, directly addressing the insider misuse and external re-identification threats outlined in the threat model.
Whereas anonymization masks identifiers after data collection, synthetic generation removes reliance on raw PHI altogether, eliminating exposure points before they arise.
By simulating clinical distributions from learned or rule-based patterns, it dismantles the economic and operational incentives for insiders who might leak raw EHR exports and renders external ransomware operators unable to monetize stolen backups or staging snapshots \cite{reference49, a7}.
Because synthetic records contain no true identifiers or one-to-one mappings to real patients, they nullify quasi-identifier risks and schema-specific leak paths that low- and moderate-resource insiders can trigger during data preparation \cite{breference1, reference94}.
For high-resource external adversaries, the value of exfiltrated data collapses; synthetic ETL snapshots or exported samples cannot be used for re-identification, blackmail, or population linkage \cite{stadler2022synthetic, narayanan2008robust}.

\textbf{Rule-based methods:}
Rule-driven and simulation approaches generate synthetic data from explicit clinical logic or population statistics, providing auditable and deterministic privacy guarantees.
Monte Carlo and discrete-event models replicate hospital workflows—admissions, lab requests, and comorbidities, without using real patient identifiers \cite{reference32, reference34}.
For instance, the Dismed dataset randomized annotated entities using biomedical ontologies to maintain diagnostic structure while erasing sensitive identifiers \cite{reference21, reference38}.
Such systems are particularly effective against insider leakage because their generation logic never touches PHI; even if preprocessing pipelines or exports are compromised, the data itself is synthetic and devoid of re-identifiable features.

\textbf{Using LLMs:}
Learning-based methods extend this protection to complex multimodal data such as radiology reports, clinical narratives, and tabular EHRs.
Generative networks (GANs, VAEs, and diffusion models) and LLMs capture nonlinear dependencies across diagnoses, medications, and outcomes \cite{reference30, reference31}.
LLM-based generators—such as GPT-4 and LLaMA 3.1-70B, have been used to produce synthetic case studies and comorbidity graphs for models like ComLLM and Asclepius, supporting clinical QA and CDS tasks without direct patient data access \cite{reference36, reference37, reference39}.
This approach directly mitigates the text-based leakage vulnerabilities identified in the threat analysis, where identifiers persist in unstructured notes or debugging logs \cite{breference3, reference94}.
By inserting a synthetic layer between hospital data and LLM pipelines, healthcare institutions convert high-risk preprocessing workflows into auditable, privacy-preserving sandboxes.

\textcolor{blue}{\textbf{Takeaway.}} 
Synthetic data generation directly mitigates the vulnerabilities identified in the preprocessing threat model by removing dependence on \textit{raw PHI} before training. 
By replacing true records with statistically valid but fictitious samples, it disrupts the \textit{economic and operational incentives} of low- and moderate-resource insiders who rely on \textit{understanding of EHR schemas and identifiers} and \textit{access to staging data} to exploit \textit{quasi-identifiers} or export unmasked EHRs. 
At the same time, synthetic data eliminates the informational value that high-resource external adversaries derive from \textit{auxiliary datasets}, rendering \textit{re-identification}, \textit{linkage attacks}, and exploitation of \textit{EHR exports} or \textit{staging leaks} ineffective. 
By removing the real identifiers that make these vulnerabilities actionable, synthetic data closes the same \textit{attack surfaces} described in the threat model and reshapes the preprocessing pipeline into a controlled environment where both insider misuse and external infiltration have substantially reduced impact.

\textcolor{violet}{\textbf{Limitation.} 
While synthetic data generation reduces dependence on real PHI, it does not eliminate preprocessing vulnerabilities.\textit{(1) Quasi-identifiers:} generators may inadvertently reproduce statistical or quasi-identifier patterns, leaving the system exposed to the same re-identification and linkage risks exploited by high-resource adversaries with strong prior knowledge and auxiliary records.\textit{(2) Weak access controls: }synthetic pipelines remain vulnerable to data poisoning introduced before generation, allowing moderate-resource insiders or malicious collaborators to manipulate raw PHI and exploit their access privileges. \textit{(3) Data leakage via logs: }if generation quality is inconsistent or insufficiently audited, synthetic datasets may recreate structural cues that allow adversaries to exploit previously identified attack surfaces such as EHR exports, staging leaks, or API exposures. Thus, while synthetic generation reduces reliance on real PHI, its effectiveness depends on strict governance and validation to prevent threat-model vulnerabilities from re-emerging through statistical artifacts.}

\subsubsection{Differential Privacy (DP)} provides a formal defense against the inference and memorization risks that persist in preprocessing, particularly when partial identifiers or structured embeddings are exposed.
Unlike anonymization or synthetic generation, DP offers quantifiable privacy guarantees by injecting calibrated noise into data values, gradients, or outputs \cite{reference28}.
This directly addresses the moderate-resource insiders and external attackers identified in the threat model, those capable of inspecting gradients, intermediate logs, or fine-tuning checkpoints \cite{breference3, reference94}.
By applying Gaussian or Laplacian noise during aggregation or embedding generation, DP ensures that individual patient records remain statistically indistinguishable, even to privileged ETL engineers or model developers \cite{reference46, reference61}.
Healthcare frameworks such as DisLLM, MedMCQ, and PubMedQA demonstrate how DP-SGD and local DP protect sensitive EHR and clinical text embeddings while retaining acceptable task accuracy \cite{reference60, reference61, reference62}.
Local DP, used in resource-limited hospital preprocessing, adds noise directly at the feature extraction stage, obscuring PHI before it reaches shared systems, while centralized DP reduces cross-phase linkage between preprocessing and downstream fine-tuning \cite{reference64}.
 Beyond strict DP, stochastic embedding methods such as NEFtune and SHADE-AD \cite{reference57, reference58} introduce non-determinism to reduce overfitting and memorization.Although lacking formal guarantees, these strategies complement DP by further diminishing the information advantage of internal or external adversaries during preprocessing.\newline
\indent \textcolor{blue}{\textbf{Takeaway.}} DP directly mitigates the inference and reconstruction risks described in the preprocessing threat model by protecting against the misuse of \textit{gradients}, \textit{intermediate logs}, and \textit{structured embeddings}. By adding calibrated noise during feature extraction or aggregation, DP reduces the ability of moderate-resource insiders, who have access to \textit{debugging logs}, \textit{staging data}, or \textit{model checkpoints}, to recover patient attributes or exploit \textit{quasi-identifiers}. The same noise also weakens external adversaries who rely on auxiliary datasets to perform \textit{re-identification} or \textit{record reconstruction}. In this way, DP converts preprocessing from a trust-dependent stage into one that enforces quantifiable privacy guarantees, directly addressing the vulnerabilities associated with linkage attacks and log leakage identified in the threat model.
\indent \textcolor{violet}{\textbf{Limitation.} 
Despite its benefits, DP does not address all vulnerabilities in the preprocessing threat model. \textit{(1) Weak access controls/APIs:} DP cannot stop data poisoning, since adversaries can inject or manipulate raw clinical records before noise is applied, exploiting insecure ETL workflows.
\textit{(2) Quasi-identifiers:} the privacy–utility trade-off can distort sensitive fields, degrading fidelity in downstream clinical tasks while leaving partial re-identification risks. \textit{(3) Data leakage via logs:} DP’s effectiveness depends on correct configuration; mis-set noise budgets or implementation flaws can still expose PHI through debug traces or unsecured system logs. Thus, while DP mitigates re-identification and log exposure, it cannot fully prevent poisoning or misconfiguration exploits in preprocessing.}

\begin{tcolorbox}[
  colback=gray!5,
  colframe=gray!20,
  boxrule=0.3pt,
  arc=1mm,
  left=2mm,
  right=2mm,
  top=1mm,
  bottom=1mm,
  enhanced,
  breakable,
  before skip=4pt,
  after skip=4pt
]
\textbf{Recommendation 1:}
The preprocessing phase exposes the most sensitive attack surface of the healthcare LLM pipeline, where adversaries exploit raw identifiers, staging artifacts, and weak system configurations. To mitigate these vulnerabilities, defenses must directly align with the threat model.
\textbf{(1) Quasi-identifiers Exposure.} Implement task-specific anonymization combined with differential privacy audits to remove or mask high-risk identifiers before ETL execution. This reduces re-identification and linkage attacks that exploit demographic and clinical quasi-identifiers using prior knowledge and auxiliary datasets
\textbf{(2) Data leakage via logs.}Adopt privacy-aware logging frameworks with automatic redaction, hashed identifiers, and secure audit retention policies. This ensures ETL traces and debugging outputs cannot expose PHI through intermediate tables or system logs accessible to insiders or compromised systems.
\textbf{(3) Weak access controls and API exposures.} Deploy role-based access control (RBAC) and zero-trust authentication across preprocessing nodes, ensuring that only verified users and processes can query or export data. Integrate secure API gateways and encryption-in-use (TEE) mechanisms to protect intermediate exports and prevent unauthorized dataset retrieval
\textbf{(4) Data poisoning.} Incorporate data integrity validation and schema-constrained ingestion to detect anomalous or manipulated records injected before anonymization.
Combine this with input provenance tracking and hash-based cross-validation to ensure the authenticity of clinical feeds (HL7/FHIR) before preprocessing pipelines ingest 
\end{tcolorbox}

Table \ref{tab:preprocessing-taxonomy} summarizes the preprocessing-phase taxonomy, consolidating internal/external actors, attack surfaces, defenses, limitations, and actionable recommendations derived from this section.

\section{Federated Fine-Tuning Phase} 
\label{sec:fine_tuning}
\vspace{-0.3em}
Building on the vulnerabilities introduced during preprocessing, the fine-tuning phase exposes a new class of distributed risks: even when raw PHI is masked, residual artifacts, cohort structures, and poisoned records propagate into gradients and client updates across federated hospitals. We focus specifically on federated fine-tuning because, in healthcare, centralized fine-tuning is rarely feasible in cross-institutional raw-data pooling violates HIPAA/GDPR, institutional policies, and multi-site data-sharing agreements \cite{reference64, reference05, breference1, 2025dataprivacy}. As a result, federated fine-tuning is not an architectural preference but the only legally and operationally viable mechanism for adapting LLMs to clinical data in real deployments \cite{paper06, reference12, paper016, paper017}. This makes FL the principal attack surface and hence the appropriate scope for a phase-aware SoK.

Fine-tuning LLMs in healthcare enables models to adapt to domain-specific terminology and tasks, improving accuracy and clinical relevance. However, fine-tuning on sensitive, institution-specific datasets introduces distinct privacy risks. While techniques such as Differential Privacy (DP) and adapter tuning offer partial protection \cite{paper013}, the majority of practical deployments and research indicate that Federated Learning (FL) is the most comprehensive privacy-preserving framework for regulated healthcare environments \cite{paper01, paper03, paper016}. In this phase, we outline the threat landscape specific to FL, focusing on the privacy challenges arising from distributed training infrastructures connecting hospitals.

\subsection{Threat Model}
\vspace{-0.3em}
In the fine-tuning phase, threats arise from adversaries, internal or external, who interact with the FL infrastructure connecting hospitals. Unlike the preprocessing stage, where unmasked raw data is exposed, here the primary privacy risks emerge from access to \textit{local gradients}, \textit{client updates}, and \textit{communication buffers}. These artifacts implicitly encode sensitive clinical attributes and thus create a distinct attack surface.

\textbf{Internal adversaries} pose the most operationally impactful risks because they operate inside trusted hospital boundaries and interact directly with local training workflows:
\textbf{(1)} Clinicians or annotators may inadvertently expose local patient records while validating model predictions or providing feedback on draft outputs \cite{reference84}.
\textbf{(2)} ML engineers and data scientists routinely inspect gradient snapshots or checkpoints during debugging, unintentionally accessing encoded EHR information \cite{reference16}.
\textbf{(3)} System administrators and IT staff managing synchronization or storage systems have privileged visibility into local model caches, communication buffers, or server logs, making silent extraction or export of update data feasible \cite{paper03}.
These internal actions rarely trigger alarms because they occur within expected operational workflows.

\textbf{External adversaries} target the distributed nature of FL and the heterogeneity of hospital networks:
\textbf{(1)} Network intruders intercept parameter updates or partial gradients in transit, enabling reconstruction of patient details via gradient inversion \cite{paper015}.
\textbf{(2)} Corporate or state-linked actors inject malicious updates to bias shared clinical models, impairing diagnostic consistency across sites \cite{paper04}.
\textbf{(3)} Ransomware and extortion groups target aggregation servers to exfiltrate multi-institutional model checkpoints encoding sensitive patterns across cohorts \cite{paper016}.
These adversaries are driven by financial, strategic, or competitive motives and exploit FL’s distributed communication as a large-scale leakage vector.

\subsubsection{Prior Knowledge and Capability Gradient}
Adversaries in this phase vary in both domain knowledge and computational capacity, which determine how effectively they can exploit gradients, checkpoints, or update buffers:
\textbf{(1)} Low-resource insiders (e.g., clinicians, annotators) understand how patient-level features are represented in model behavior and may unintentionally expose encoded EHR details during validation or feedback. \textbf{(2)} Moderate-resource insiders (data scientists, ML engineers) possess deeper knowledge of model architectures and storage layouts, knowing where gradient files, synchronization buffers, or checkpoints are stored, allowing silent access or export during debugging or maintenance \cite{paper03}. \textbf{(3)} High-resource external adversaries, such as coordinated ransomware groups or corporate actors, leverage prior knowledge of biomedical embeddings and auxiliary EHR datasets to align intercepted gradients with public checkpoints, reconstructing rare conditions or site-specific vocabulary \cite{paper015,reference28}.

This knowledge gradient corresponds directly to their operational capabilities:
\textbf{(1)} Low-resource insiders leak updates through logs and validation workflows;
\textbf{(2)} Moderate-resource insiders or externals perform gradient inversion and membership inference using intercepted traffic or open biomedical corpora \cite{paper04,paper08}; and
\textbf{(3)} High-resource actors conduct model alignment or poisoning across hospitals, exploiting the distributed update-sharing and aggregation layers \cite{reference16,paper016}.
Together, these escalating capabilities explain how prior knowledge, from local system familiarity to cross-institutional data access, translates directly into exploitation of fine-tuning vulnerabilities such as gradient leakage, client update interception, and poisoning within federated healthcare networks.
\subsubsection{Attack Surface Vulnerabilities and Enabled Attacks}
Distinct attack surface vulnerabilities emerge during fine-tuning of healthcare LLMs, each tied to the nature of clinical datasets and federated infrastructures.

\begin{itemize}
    \item \textbf{Model gradient leakage:} During fine-tuning, gradients or model weights stored locally may inadvertently memorize sensitive clinical data such as structured EHR variables (e.g., blood pressure, cholesterol levels, ICD-10 codes) and unstructured discharge narratives. Even partial leakage can expose individual patient trajectories, particularly for rare diseases recorded in small data sets. For example, in cardiology, the leakage of gradients can reveal critical patient outcomes like ejection fraction or surgical interventions—easily re-identifiable in small cohorts \cite{paper0034, reference16}
    \item \textbf{Client update leakage:} In federated fine-tuning, hospitals share model updates instead of raw client data.  However, these updates still carry sensitive EHR patterns, such as lab panels, clinical notes, or medication histories, which can still be leaked. Adversaries may use gradient inversion to reconstruct detailed narratives (e.g., ICU notes, pathology reports). Moreover, poisoning attacks may skew diagnoses, such as for diabetes or heart failure\cite{paper016,reference12}. This makes federated updates particularly sensitive, as partial feature leakage can compromise longitudinal EHR timelines.
    \item \textbf{Communication channel interception:} During model synchronization, adversaries who intercept updates can recover latent embeddings associated with sensitive clinical data such as EHR-derived phenotypes,  lab trajectories for oncology patients, or cardiology monitoring logs. In multilingual hospital systems, intercepted updates can also expose doctor–patient transcripts integrated with EHRs, potentially revealing sensitive lifestyle or family history details tied to EHRs\cite{paper04,paper017}.
    \item \textbf{Misconfigured audit and logging systems:} Many logging mechanisms often capture sensitive traces present in model training data, including PHI tokens from EHR fields such as medication lists, allergies, or comorbidities. If these logs are misconfigured or improperly secured, adversaries can  gain access to sensitive metadata such as genetic risk markers or rare disease cohorts, which  could then be cross-linked with external datasets\cite{reference14,reference10}.
    
\end{itemize}
These vulnerabilities directly enable several concrete attack types in federated fine-tuning: 
\textbf{(1) Gradient inversion attacks}, where adversaries reconstruct sensitive clinical details, such as ICU notes, pathology descriptions, or lab trajectories from leaked or intercepted gradients. 
\textbf{(2) Membership inference attacks}, which allow adversaries to determine whether a specific patient’s records contributed to the fine-tuning process, particularly for rare diseases or small-site cohorts. 
\textbf{(3) Update leakage attacks}, where model updates shared across hospitals reveal structured or unstructured EHR patterns embedded in weight deltas or optimizer states. 
\textbf{(4) Model poisoning and backdoor attacks}, where malicious insiders or externals inject crafted updates that bias diagnosis-related outputs or embed hidden triggers into clinical prediction tasks. 
\textbf{(5) Communication-layer interception attacks}, where man-in-the-middle adversaries extract latent representations from synchronization streams exchanged between hospitals and the central aggregator. 

\subsection{Privacy-Preserving Defenses}

This section reviews core privacy defenses integrated into federated learning (FL) fine-tuning workflows covering client-side, secure update sharing, and communication safeguards followed by limitations and recommendations.

\subsubsection{Client Side}
These cover the privacy-preserving techniques employed at the client side during training in the FL setup.

\textbf{Differential Privacy (DP)} adds calibrated noise to model updates before leaving the hospital, preventing patient-level re-identification in sensitive datasets such as MIMIC-IV ICU notes or oncology records \cite{reference64}. DP-LoRA applies noise only to adapter layers, preserving accuracy \cite{paper07}, while selective DP targets the most sensitive parameters \cite{paper01,paper010}. Despite potential signal loss in rare disease cohorts \cite{paper0034}, DP remains a practical safeguard for HIPAA/GDPR compliance and multi-hospital training \cite{reference70}.

\textbf{Secure Multi-Party Computation (SMPC)} enables hospitals to train jointly without sharing raw data. Each site encrypts and splits its updates, which are only usable when aggregated \cite{reference70}. This supports legally restricted studies like cancer survival or cardiovascular prediction \cite{reference10}. While secure against reconstruction, SMPC’s high compute cost limits real-time applications \cite{reference64}, though it remains vital for regulated hospital consortia \cite{paper016}.

\textbf{Split Learning (SL) } 
partitions the model between client and server—local layers process raw data, and only intermediate features are shared \cite{reference70,reference64}. This protects imaging and textual data (e.g., ChestX-ray14, ECG datasets) while supporting multimodal tasks. Though intermediate features may leak partial identity traces, SL’s lightweight setup enables participation from smaller hospitals with limited infrastructure.

\textbf{Randomized Low-Rank Adaptation (LoRA)} fine-tunes only low-rank parameters with added randomization, hindering inversion attacks on sensitive text (e.g., oncology notes, psychiatric transcripts) \cite{paper07,paper010}. It offers better accuracy than DP and can be combined with it for stronger protection \cite{paper05}. Its low compute footprint makes it ideal for hospitals with limited GPUs, balancing privacy, efficiency, and performance.

\textbf{Quantization} 
reduces weight precision (e.g., 32-bit → 8/4-bit), limiting exploitable detail while cutting memory and bandwidth needs \cite{reference70,reference10}. Effective on structured medical data, quantized models maintain accuracy and resist gradient inversion \cite{paper014}. When paired with DP or LoRA, it forms a layered defense which is resource-efficient and scalable for diverse healthcare networks \cite{paper01,paper09,paper016}.

\textcolor{blue}{\textbf{Takeaway.}} Client-side mechanisms directly address privacy risks arising from local training gradients and intermediate checkpoints.
\textit{DP} mitigates memorization and gradient inversion by adding calibrated noise, reducing the ability of internal or external adversaries to reconstruct structured or narrative EHR data from local model states.
\textit{SMPC} encrypts local gradients before aggregation, preventing adversaries from extracting embeddings or conversational features from intercepted updates.
SL minimizes raw data transmission by keeping sensitive features (e.g., imaging, ECG traces, pathology text) local and sharing only intermediate activations.
\textit{Randomized LoRA} introduces stochasticity in parameter updates, weakening the consistency needed for gradient correlation or poisoning.
\textit{Quantization} further obscures precise gradient values, reducing leakage in logs and audit traces while lowering communication overheads.
Together, these defenses collectively address gradient leakage, poisoning, and metadata exposure across distributed clients.

\textcolor{violet}{\textbf{Limitation.} 
Despite reducing risks, client-side defenses still leave healthcare FL systems exposed across identified attack surfaces.
\textit{(1) Model Gradient Leakage:} DP reduces signal strength but the privacy–utility trade-off can suppress rare-disease features while leaving residual exposure in local artifacts (e.g., verbose checkpoints or cached tensors), enabling gradient inversion and membership inference on small cohorts.
\textit{(2) Client Update Leakage:} SMPC hides individual updates from the aggregator but does not protect against leakage at the endpoints (misconfigured clients/servers, optimizer state dumps) and does not prevent update reconstruction from artifacts captured outside the secure aggregation step; it also does not address poisoning/backdoor manipulation of valid shares.
\textit{(3) Communication Channel Interception:} Split learning keeps raw inputs local, yet exchanged intermediate representations and synchronization streams remain linkable; advanced adversaries can perform re-identification/linkage on intercepted communication buffers or server-visible activations.
\textit{(4) Misconfigured Audits:} Randomized LoRA or quantization does not sanitize logs; if audit systems record local gradients, client updates, or intermediate tensors, correlation across rounds/sites can still expose sensitive dialogues or discharge notes.
These limitations show that current defenses partially address risks like memorization and poisoning, leaving vulnerabilities for adversaries to exploit in federated healthcare training.}

\begin{table*}[t]
\centering
\caption{Federated Fine-Tuning Phase — Summary of Threats, Defenses, Limitations, and Recommendations.}
\label{tab:federated-taxonomy}

\renewcommand{\arraystretch}{1.25}
\footnotesize
\setlength{\tabcolsep}{3.8pt}

\begin{tabularx}{\textwidth}{|>{\raggedright\arraybackslash}p{4cm}|
                                >{\raggedright\arraybackslash}p{4cm}|
                                >{\raggedright\arraybackslash}p{4cm}|
                                >{\raggedright\arraybackslash}p{4cm}|}
\hline
\rowcolor[HTML]{E8E8E8}
\textbf{Threat Model} & \textbf{Privacy-preserving Defenses} & \textbf{Limitations} & \textbf{Recommendations} \\
\hline

\begin{itemize}[leftmargin=*,nosep]
    \item \textbf{Internal adversaries:} clinicians/annotators (validation leaks); ML engineers (gradients, checkpoints); system admins (caches, sync buffers).
    \item \textbf{External adversaries:} network intruders, ransomware groups, state-linked or corporate actors.
    \item \textbf{Capabilities:} gradient inversion, membership inference, poisoning/backdoors, alignment with auxiliary corpora, client-update reconstruction.
    \item \textbf{Key Vulnerabilities:} local gradients, client updates, optimizer states, communication buffers, synchronization traffic, misconfigured logs/audits.
\end{itemize}
&
\begin{itemize}[leftmargin=*,nosep]
    \item \textbf{Client-side:} DP-SGD, DP-LoRA, selective/local DP; SMPC for encrypted/split updates; Split Learning (local early layers); randomized LoRA; quantization for low precision (8/4-bit).
    \item \textbf{Client update sharing:} Secure Aggregation (masking, LoRA-only); Weight-Delta sharing (adapter updates only); Blockchain for integrity, tamper-evidence, and unlearning.
    \item \textbf{Communication channel:} Adapter-based compression (LoRA), few-shot learning (small gradients), and RTIR (lightweight reasoning fine-tuning).
\end{itemize}
&
\begin{itemize}[leftmargin=*,nosep]
    \item \textbf{Model gradient leakage:} DP degrades rare-disease fidelity; cached tensors persist; Split Learning leaks intermediate activations.
    \item \textbf{Client update leakage:} SMPC protects only at aggregation; poisoning and backdoors persist; deltas reconstructable across rounds.
    \item \textbf{Communication interception:} LoRA lowers size but not local leakage; few-shot gradients reconstructable; RTIR depends on secure retrieval.
    \item \textbf{Misconfigured logs/audits:} logs capture gradients/activations; blockchain metadata can reveal institutional identifiers.
\end{itemize}
&
\begin{itemize}[leftmargin=*,nosep]
    \item \textbf{Gradient leakage:} use adaptive DP with randomized LoRA to reduce inversion and membership inference.
    \item \textbf{Update leakage:} combine SMPC with clipping and anomaly detection; selectively freeze sensitive layers.
    \item \textbf{Communication interception:} adopt authenticated encryption, quantized aggregation, and TEEs; pair LoRA with DP.
    \item \textbf{Audit exposure:} use privacy-aware provenance; minimize logged gradients; obfuscate ledger metadata.
    \item \textbf{Combined mitigation:} integrate quantization + adaptive DP + randomized LoRA for fidelity, low bandwidth, and poisoning resistance.
\end{itemize}
\\
\hline
\end{tabularx}
\label{tab:tab2}
\end{table*}

\subsubsection{Client Update Sharing and Secure Aggregation}
This stage targets vulnerabilities in update transmission and aggregation, where even privacy-preserving local training can expose sensitive patterns. Because the aggregator is a central risk point, mechanisms such as secure aggregation and blockchain-based FL are critical in healthcare to maintain both confidentiality and institutional accountability.

\textbf{Secure Aggregation} ensures that hospital updates remain confidential during cross-silo fine-tuning. Each client masks its gradients or adapter updates with random values; only the summed result reveals the true aggregate \cite{paper016,paper015,paper010}. This prevents reconstruction of sensitive features such as medication histories or lab trajectories. To reduce overhead, recent work masks only LoRA adapter weights instead of full gradients \cite{paper07,paper01}, while anomaly detection helps flag poisoned updates targeting rare-disease cohorts \cite{reference10}. Though computationally demanding for small hospitals \cite{reference70}, secure aggregation remains essential in federated healthcare networks, balancing privacy, compliance, and trust.

\textbf{Weight Delta Sharing.} transmits only parameter differences between local and global models, minimizing exposure of raw patient data \cite{paper014}. This approach improves efficiency for EHR-based LLMs such as MIMIC-IV discharge models \cite{reference57}. LoRA-only delta sharing compresses updates into low-rank matrices for multilingual hospital networks \cite{paper04}, while selective layer freezing focuses on clinically relevant upper layers \cite{paper0034,paper03}. However, repeated deltas can leak sensitive trends; combining delta sharing with DP or gradient clipping mitigates this risk \cite{paper010}.

\textbf{Blockchain for Update Integrity and Unlearning.}
provides traceability and auditability by recording encrypted update hashes, timestamps, and institutional IDs \cite{paper08}. This deters tampering and enables federated unlearning, removing a client’s contributions without full retraining \cite{paper08,paper05}. Provenance records further enhance institutional trust in federated healthcare deployments \cite{reference70}. While added infrastructure and latency pose challenges, blockchain remains valuable for high-stakes domains such as oncology or ICU consortia \cite{paper03,paper01}.

 \textcolor{blue}{\textbf{Takeaway.}}
At the aggregation layer, update-sharing mechanisms mitigate vulnerabilities associated with cross-site synchronization and multi-round leakage. \textit{Secure aggregation} limits client update leakage at the aggregator and weakens gradient inversion and membership inference on per-site contributions by preventing inspection of individual updates. \textit{Weight-delta sharing }narrows exposure but must be paired with DP or clipping to resist multi-round leakage and subsequent update reconstruction. \textit{Blockchain-based provenance} strengthens protection against misconfigured audit and logging systems and helps surface poisoning/backdoors by providing tamper-evident integrity and traceability for updates. Quantized aggregation alleviates communication-channel bottlenecks by compressing updates, though aggressive compression can affect rare-disease fidelity and bias.

\textcolor{violet}{\textbf{Limitation.} Update-sharing defenses reduce some risks but still leave critical gaps when mapped to the attack surfaces.
\textit{(1) Client update leakage:} Secure aggregation conceals individual hospital updates at the aggregator, but it does not prevent poisoning/backdoors adversaries can inject crafted, valid-looking shares, and endpoint artifacts (caches, optimizer states) can still leak per-site information.
\textit{(2) Multi-round leakage (client update leakage):} Weight-delta sharing lowers bandwidth per round, yet deltas accumulated across rounds enable reconstruction and membership inference over pathology or ICU trajectories, sustaining the multi-round leakage surface without additional DP or clipping.
\textit{(3) Misconfigured audit and logging systems:} Blockchain provides tamper evidence, not confidentiality; if ledger or audit metadata (institution IDs, timestamps, cohort counts) are exposed or logs are unsanitized, cross-linking and profiling of rare-disease cohorts remain possible on the audit surface.
\textit{(4) Communication channel interception:} Quantized aggregation compresses updates but does not secure transport or endpoints; intercepted compressed streams can still be analyzed, and aggressive rounding degrades rare-disease fidelity, opening room for biased predictions in oncology or cardiology tasks.}

\subsubsection{Communication Channel}
As LLMs integrate into federated learning (FL), communication overhead becomes a major constraint, distinct from classical FL threats. In healthcare, transmitting full model updates is impractical; hence, reducing transmission size and frequency is crucial for scalability, privacy, and performance.

\textbf{Adapter-based Compression} such as LoRA minimizes communication by updating only low-rank adapter parameters rather than full model weights \cite{paper07}. This allows hospitals to share compact updates while keeping billions of base parameters frozen, cutting bandwidth use \cite{paper014,paper012,paper05}. Studies such as Med42 confirm LoRA-based fine-tuning maintains accuracy for oncology classification and discharge summary generation while significantly reducing parameter size \cite{paper0034}. LoRA-only updates also enable multilingual medical transcript training across silos \cite{paper04} and limit PHI exposure since fewer parameters leave the institution \cite{paper010,paper06}. Additionally, adapter-level sharing supports selective unlearning, removing hospital-specific contributions without retraining \cite{paper03}. However, LoRA may underperform on deep multimodal tasks like Clipsyntel summarization that require richer representations \cite{reference15}.

\textbf{Few-shot Learning} reduces communication by fine-tuning on limited local samples (10–50), yielding sparse, lightweight gradients \cite{paper09,paper017}. Applied to ICU discharge notes and oncology reports, this method transmits only task-relevant updates while maintaining accuracy \cite{reference57}. Few-shot FL further supports multilingual transcripts for hospitals with small datasets \cite{paper04,paper011}. Smaller updates inherently reduce inversion risk \cite{paper010}, and combining few-shot training with DP or adapter tuning enhances stability \cite{paper0035}. Still, performance drops in multimodal settings like Clipsyntel-based question summarization, which demands deeper contextual learning \cite{reference15}.

\textbf{Real-Time Information Retrieval (RTIR)} decouples knowledge from model weights, retrieving external data during inference instead of embedding it in updates \cite{paper01,paper04}. In healthcare, this allows querying clinical knowledge bases for oncology guidelines, multilingual symptom data, or ICU protocols without transmitting sensitive gradients. Hospitals then fine-tune only lightweight reasoning layers, dramatically shrinking update size and preventing patient data memorization \cite{reference14}. RTIR integrated with adapter tuning supports efficient and privacy-preserving federated pipelines while sustaining diagnostic accuracy \cite{paper02}. The trade-off lies in reliance on secure retrieval infrastructure, which may challenge smaller hospitals, and its limited autonomy in decision-making. Nonetheless, RTIR offers a dynamic, privacy-resilient strategy for fast-evolving healthcare domains.\newline
\indent \textcolor{blue}{\textbf{Takeaway.}} Channel-level defenses map to the vulnerabilities identified in the threat model. Adapter-based Compression (LoRA) reduces \textit{bandwidth usage} and narrows client update leakage by transmitting only adapter updates, weakening opportunities for gradient inversion and membership inference over communication buffers. Few-shot Learning yields sparse, smaller updates from limited local examples, further reducing the signal available for update reconstruction on sensitive EHR trajectories and pathology timelines and lowering exposure during \textit{communication channel interception}. RTIR decouples knowledge from model weights so entire clinical histories are not embedded in checkpoints, shrinking the footprint of client updates and the attack surface for interception and downstream leakage. Collectively, these methods target \textit{interception, leakage, and channel vulnerabilities}, reducing adversarial opportunities in federated healthcare synchronization streams.

\textcolor{violet}{\textbf{Limitation.} Communication-efficient methods reduce bandwidth but still leave key threat surfaces exposed in healthcare FL.
\textit{(1) Model gradient leakage:} Adapter-based compression (LoRA) restricts parameter sharing but does not sanitize local artifacts; reduced capacity can weaken rare-disease fidelity while leaving residual signal for gradient inversion and membership inference on the gradient/memorization leakage surface.
\textit{(2) Client update leakage:} Few-shot learning minimizes exchanged updates, yet sparse gradients can omit subtle clinical signals (e.g., drug–drug interactions) and remain vulnerable to multi-round reconstruction and poisoning/backdoors, sustaining exposure on the client update surface and risking biased predictions for under-represented tasks.
\textit{(3) Communication channel interception:} RTIR avoids embedding PHI in weights, but insecure retrieval paths and cached queries keep the communication buffers susceptible to interception and cross-linking (e.g., transcripts, sensitive queries for HIV or psychiatric histories).}

\begin{tcolorbox}[
  colback=gray!5,
  colframe=gray!20,
  boxrule=0.3pt,
  arc=1mm,
  left=2mm,
  right=2mm,
  top=1mm,
  bottom=1mm,
  enhanced,
  breakable,
  before skip=4pt,
  after skip=4pt
]
\textbf{Recommendation 2:} To close the defensive gaps identified in the threat model, future federated healthcare systems should employ layered, phase-specific defenses explicitly mapped to the key vulnerabilities. \textbf{(1) Model gradient leakage.} Adaptive DP should be combined with Randomized LoRA to mitigate gradient inversion and memorization risks while preserving fidelity for rare-disease or small-cohort training. This pairing directly strengthens protection against internal engineers or external interceptors who exploit gradient sensitivity.\textbf{(2) Client update leakage.} Augment Secure Aggregation and Weight-Delta Sharing with gradient clipping and anomaly detection to detect poisoning and reconstruction attempts before aggregation. These measures counter moderate-resource adversaries who manipulate or infer sensitive EHR patterns from multi-round updates. \textbf{(3) Communication channel interception.} Deploy hybrid cryptographic schemes combining lightweight SMPC and TEE-based aggregation to encrypt and isolate update streams while maintaining acceptable latency. This limits interception and timing analysis by external or cross-institutional adversaries during synchronization over WAN links. \textbf{(4) Misconfigured audits/logging.} Integrate blockchain-based provenance tracking with metadata obfuscation and differentially private logging to ensure traceability without exposing participation frequency or institutional identifiers. This reduces internal leakage risks from logs and external deanonymization via audit metadata.
\end{tcolorbox}

Table \ref{tab:federated-taxonomy} provides the fine-tuning-phase taxonomy, capturing the federated update pathway, client/server adversaries, surface-specific threats, and the layered defenses discussed here.
\vspace{-1.2em}
\section{Inference Phase}
\label{sec:inference}
The vulnerabilities embedded during preprocessing and compounded through federated fine-tuning surface most clearly during inference, where PHI can leak through prompts, hidden states, caches, and outputs even when model parameters remain protected.
In healthcare, LLM inferencing is the process where a trained LLM uses patient data to generate predictions, diagnoses, and personalized treatment recommendations in real time. However, this phase poses a distinct threat in LLM deployments, as clinical inputs can leak through outputs, embeddings, or memory without exposing model parameters. Protecting patient data during inference is essential for regulatory compliance and ethical AI. 

\subsection{Threat Model}
\vspace{-0.3em}
Inference in healthcare LLM deployments introduces risks that differ from training. Here, PHI can leak via prompts, embeddings/hidden states, KV caches, and returned text, even when model weights and training data are protected. This matters for real clinical tools like discharge-note assistants, triage chatbots, radiology and pathology summarizers, and multilingual transcript systems—often deployed on hybrid edge–cloud stacks under tight latency constraints \cite{reference16,reference14,reference84,reference15,paper04}.

\subsubsection{Attacker Landscape and Incentives}Inference introduces distinct adversarial incentives because sensitive information appears in runtime artifacts, \textit{not} in training data or gradients. These artifacts persist in logs, caches, monitoring traces, and activation buffers maintained across large, distributed clinical infrastructures.

\textbf{Internal adversaries} pose the most persistent risk, as they interact directly with operational LLM systems deployed inside hospitals:
\textbf{(1)} IT operators, SRE teams, and platform engineers often access telemetry dashboards, runtime logs, and GPU memory during debugging. Because multi-turn systems maintain \textit{KV caches} across generations, these memory snapshots may reveal full conversations, including telemedicine notes or oncology assessments \cite{paper033,paper034}.
\textbf{(2)} Clinicians and data scientists frequently input full EHR excerpts (e.g., ICU timelines, staging notes, surgical histories) into prompts during validation or A/B testing. These prompts can resurface in logging, error traces, or monitoring pipelines \cite{paper026,reference14}.
\textbf{(3)} Teams operating domain-specific inference pipelines, such as digital pathology or endocrine-cancer extraction systems, may store or inspect intermediate embeddings during performance tuning, inadvertently exposing PHI embedded in feature vectors \cite{reference84}.
Internal threats are especially severe because these activities occur under legitimate operational workflows and rarely trigger intrusion detection.
External adversaries

\textbf{External adversaries} raise threats to exploit LLM deployment APIs, cross-hospital WAN links, or cloud inference infrastructure:
\textbf{(1)} API-based attackers can perform \textit{user inference attacks}, probing outputs to determine whether the model was adapted on rare clinical cohorts (e.g., rare oncology patients), enabling deanonymization of small hospitals \cite{paper019,paper022}.
\textbf{(2)} Cloud-side or semi-trusted platform adversaries can apply \textit{black-box inversion}, reconstructing sensitive spans from intermediate activations or returned outputs \cite{paper024,paper032}.
\textbf{(3)} Network adversaries exploit WAN traffic between local hospitals and cloud inference nodes; timing and packet-size patterns can reveal PHI density or prompt structure, even when content is encrypted \cite{paper027}.
\textbf{(4)} Distributed inference architectures especially those using adapter offloading or split execution expand the attack surface when activations cross trust boundaries \cite{paper029,paper025}.
These attackers operate outside institutional boundaries, motivated by financial, competitive, or strategic objectives.

\subsubsection{Prior Knowledge and Capability Gradient}
Adversaries at inference time vary in domain awareness and technical capability, shaping how effectively they exploit runtime artifacts such as prompts, KV caches, and hidden states.
\textbf{(1)} Low-resource insiders, clinicians, analysts, or IT operators understand prompt formats and may access logs or traces containing raw PHI. Their local familiarity with EHR content and dashboard telemetry enables unintentional exposure through debugging or monitoring systems \cite{paper026,paper034}.
\textbf{(2)} Moderate-capability actors, including engineers or cloud operators, know where GPU dumps, activation buffers, and inference logs are stored, allowing silent extraction or correlation of prompts, embeddings, and conversational states \cite{paper033}.
\textbf{(3)} High-resource adversaries such as external API probers or state-linked entities, possess auxiliary clinical datasets and use prior statistical knowledge to align output distributions or hidden-state signatures with real cohorts, revealing rare conditions or site-specific attributes \cite{paper019,paper022}.

This knowledge gradient translates directly into exploitability:
(1) insiders cause accidental PHI leakage via logs and KV caches;
(2) mid-level actors conduct embedding inversion or output-correlation attacks using API or infrastructure access \cite{paper024,paper028}; and
(3) high-resource adversaries perform cross-dataset alignment and timing analysis across WAN links to infer sensitive patterns or prompt structures \cite{paper027,paper029}.
\subsubsection{Attack Surface Vulnerabilities and Enabled At-
tacks}
The key vulnerabilities include:
\begin{itemize}
    \item \textbf{Prompt channel and Hidden-State Leakage:}
    Clinical prompts carry PHI such as diagnoses, medications, lab timelines, and family histories. Even when transmitted over TLS, prompts are plaintext at the service node, and simple redaction can distort clinical meaning. Hidden states store rich patterns from EHR notes or pathology text, which can be reconstructed by embedding inversion in semitrusted clouds or debug pipelines \cite{paper024,paper026,paper029,paper032}. Edge–cloud splits or adapter offloads can  leak intermediate activations if not properly protected \cite{paper025,paper029}.

    \item \textbf{KV cache leakage:}
     Multi-turn KV caches retain contextual information for efficiency. If snapshots of GPU memory or caches are exposed, entire patient dialogues, such as telemedicine or oncology sessions, can be recovered \cite{paper033,paper034}. Hybrid-cloud monitoring and scaling layers introduce additional points of exposure.

    \item \textbf{Returned text (outputs):}
    Outputs may leak PHI via over-specific recommendations (dose names, rare comorbidity patterns) or verbatim regurgitation of earlier prompts. Even with input obfuscation, outputs may re-expose sensitive facts if protections are not end-to-end\cite{paper026,paper028,reference15}.
    
    \item \textbf{Delegated / hybrid inference (edge–cloud):}
     When embeddings or adapters are computed locally and later processed in the cloud, activations and metadata traverse WAN links. Without secure partitioning or local DP, these streams can be profiled or linked to patient data \cite{paper029,paper025,paper031}. WAN conditions can also create timing side channels that leak information \cite{paper027}.
    
    \item \textbf{Restoration/meta-vector channels:}
    Pipelines that remove sensitive spans and send noised restoration vectors risk leaking protected attributes (e.g., HIV status, hereditary cancer risk) if noise schedules are mismanaged \cite{paper028}. Reusing obfuscation mappings across sessions increases this risk \cite{paper026}.
    
    \item \textbf{Operational logging and provenance:}
    LLM stacks with tracing, A/B testing, or auditing can inadvertently capture PHI in prompts, activations, or outputs. Without strict log minimization, this creates long-term leakage vulnerabilities \cite{paper034,reference64,paper02}.
\end{itemize}

These vulnerabilities enable concrete inference-time attack types highly relevant to healthcare:
\textbf{(1) Prompt reconstruction attacks}, extracting PHI from hidden states, KV caches, or traced activations.
\textbf{(2) Embedding inversion attacks}, reconstructing clinical notes, lab summaries, or pathology descriptions via inversion of hidden states or adapter outputs.
\textbf{(3) Output-based inference attacks}, inferring patient cohort membership, rare disease participation, or underlying text from over-specific model responses.
\textbf{(4) Split-inference intercept attacks}, recovering sensitive activations or adapter states crossing WAN links in hybrid edge–cloud deployments.
\textbf{(5) Restoration-vector attacks}, recovering masked identifiers or sensitive attributes from poorly noised reconstruction vectors.
\textbf{(6) Log-correlation attacks}, linking traces across clinical sessions to reconstruct longitudinal care histories.
Together, these attacks operationalize the incentives and capabilities outlined above, turning routine inference pathways into high-value exploitation channels.

\begin{table*}[t]
\centering
\caption{Inference Phase — Summary of Threats, Defenses, Limitations, and Recommendations.}
\label{tab:inference-taxonomy}

\renewcommand{\arraystretch}{1.25}
\footnotesize
\setlength{\tabcolsep}{3.8pt}

\begin{tabularx}{\textwidth}{|>{\raggedright\arraybackslash}p{4cm}|
                                >{\raggedright\arraybackslash}p{4cm}|
                                >{\raggedright\arraybackslash}p{4cm}|
                                >{\raggedright\arraybackslash}p{4cm}|}
\hline
\rowcolor[HTML]{E8E8E8}
\textbf{Threat Model} & \textbf{Privacy-preserving Defenses} & \textbf{Limitations} & \textbf{Phase-Tied Recommendations} \\
\hline

\begin{itemize}[leftmargin=*,nosep]
    \item \textbf{Internal adversaries:} IT/SRE teams with access to logs, GPU memory, KV caches; clinicians or data scientists entering full EHR prompts; teams inspecting intermediate embeddings.
    \item \textbf{External adversaries:} API-based attackers, cloud-side inversion agents, WAN interceptors, and adapter/split execution exploiters.
    \item \textbf{Capabilities:} prompt reconstruction, embedding inversion, output-based inference, cross-session correlation, KV-cache scraping, timing and metadata profiling.
    \item \textbf{Key Vulnerabilities:} prompts/inputs, hidden states, KV caches, returned text, edge–cloud activations, restoration/meta-vectors, operational logs/provenance traces.
\end{itemize}
&
\begin{itemize}[leftmargin=*,nosep]
    \item \textbf{Local DP:} DP-Forward, split-and-denoise, token perturbations (InferDPT).
    \item \textbf{Input Obfuscation/Span Removal:} PrivacyRestore, dynamic substitution/hashing, selective masking.
    \item \textbf{Cryptographic Protocols:} MPC-minimized, PermLLM (HE), GPU-accelerated FHE.
    \item \textbf{KV Cache Protection:} KV-Shield, PFID TEEs.
    \item \textbf{Model Partitioning/Edge–Cloud:} PFID local embeddings/adapters; PrivateLoRA for local adapter execution.
    \item \textbf{Federated Inference:} eFedLLM for distributed inference across institutions.
\end{itemize}
&
\begin{itemize}[leftmargin=*,nosep]
    \item \textbf{Prompt/Hidden-State Leakage:} LDP reduces fidelity; static obfuscation bypassed by cross-session correlation.
    \item \textbf{KV Cache Leakage:} relies on trusted hardware; persistent or misconfigured caches expose full dialogues.
    \item \textbf{Returned Text/Restoration:} span removal fails under repeated probing; restoration vectors reveal patient traits.
    \item \textbf{Hybrid Inference Leakage:} MPC/HE latency exposes timing channels; WAN activations profiled by network adversaries.
    \item \textbf{Operational Logging:} traces retain sensitive prompts/activations, creating persistent PHI leakage.
\end{itemize}
&
\begin{itemize}[leftmargin=*,nosep]
    \item \textbf{Prompt/Hidden-State Protection:} use span-level adaptive DP with contextual obfuscation for clinical entities.
    \item \textbf{Cross-Session Defense:} employ session-specific token randomization and dynamic obfuscation schedules.
    \item \textbf{KV Cache Security:} enforce automated expiration, TEE-based isolation, and strict memory hygiene.
    \item \textbf{Secure Hybrid Inference:} combine MPC with GPU-accelerated HE; run early layers locally before transmission.
    \item \textbf{Operational Logging/Provenance:} minimize trace capture; anonymize logs; use privacy-aware provenance tracking.
\end{itemize}
\\
\hline
\end{tabularx}
\label{tab:tab3}
\end{table*}

\subsection{Privacy-Preserving Defenses }
Inference-time attacks differ from training risks by targeting live system outputs, embeddings, and caches - often assumed non-sensitive. In healthcare, this underscores the need for holistic, privacy-by-design defenses protecting both user inputs and transient data. These mechanisms aim to prevent leakage during runtime execution, when PHI is most exposed.

\textbf{Local Differential Privacy (LDP).} introduces calibrated noise to inputs or embeddings before model access, limiting adversarial inference. DP-Forward adds noise during the forward pass to obscure subtle prompt differences \cite{paper032}, while Split-and-Denoise combines token masking with local DP to resist reconstruction \cite{paper023}. InferDPT adapts this for black-box inference with token-level perturbations \cite{paper024}. For instance, a query like “blood in stool with recurring headaches” is perturbed locally, preserving ICD-prediction utility while masking identity cues. Excessive noise, however, may weaken rare-disease fidelity.

\textbf{Input Obfuscation and Privacy Span Removal.}
selectively mask or substitute sensitive spans like names, dosages, or diagnoses before transmission. PrivacyRestore removes PHI locally, transmitting noised meta-vectors for server-side reconstruction \cite{paper028}; Instance Obfuscation hashes tokens dynamically per session \cite{paper026}. In clinical chatbots, such obfuscation ensures confidential prompts (e.g., “tested positive for HIV”) never leave the device. When paired with local DP, these lightweight methods reinforce cross-session privacy while preserving reasoning quality, ideal for telemedicine and triage.

\textbf{Cryptographic Protocols (MPC and HE).} 
protect inference over untrusted clouds. MPC-minimized secret shares only early layers to hide embeddings efficiently \cite{paper030}; PermLLM uses HE with lightweight permutations for secure attention under WAN latency \cite{paper027}; and GPU-accelerated FHE supports encrypted inference for radiology or pathology batch tasks \cite{paper020}. Though resource-intensive, such methods suit regulated domains like oncology where security outweighs delay.

\textbf{KV Cache Protection.} addresses persistent memory risks. KV-Shield permutes attention matrices to render stolen caches useless \cite{paper033}, while PFID confines cache-sensitive operations within Trusted Execution Environments (TEEs) \cite{paper029}. These safeguards prevent full-session transcript recovery in telemedicine settings with minimal latency impact, assuming secure hardware is available.

\textbf{Model Partitioning and Edge-Cloud Collaboration.} localize sensitive processing by executing early layers or adapters on secure devices. PFID runs embeddings within TEEs before delegating deeper computation \cite{paper029}, while PrivateLoRA fine-tunes and executes low-rank adapters locally \cite{paper025}. This allows mobile clinical apps to handle initial contexts (e.g., “47-year-old diabetic with chest pain”) privately before cloud inference. While device demands increase, emerging mobile accelerators make shallow inference viable.

\textbf{Federated Inference and Decentralized Deployment.} extend FL principles to inference-time privacy. eFedLLM distributes inference across multiple nodes so no single server processes full prompts \cite{paper031}. Used in oncology and multilingual telemedicine networks, each institution executes partial inference and contributes to aggregate predictions \cite{paper016}. This decentralization reduces central memory risk but introduces synchronization challenges across heterogeneous hospital systems.

 \textcolor{blue}{\textbf{Takeaway.}}
Inference-time threats arise not from raw data or gradients but from runtime artifacts, including prompts, hidden states, KV caches, intermediate activations, and returned text, where PHI may persist across sessions. Defenses such as Local DP, privacy-span removal, cryptographic inference (MPC/HE), KV-cache shielding, and edge–cloud partitioning align directly with the attack surfaces identified in the threat model. \textit{Prompt and Hidden-State Leakage:} Local DP and input obfuscation prevent direct re-identification from plaintext prompts and hidden activations, reducing the risk of embedding inversion or membership inference attacks. \textit{KV-Cache Exposure:} TEEs and KV-Shield protect multi-turn dialogue caches stored in GPU memory, securing patient–clinician transcripts in telemedicine and ICU chat scenarios. \textit{Output and Restoration Risks:} Privacy restore mechanisms remove or noise sensitive spans in returned text, limiting PHI resurfacing across prompts. \textit{Delegated/Hybrid Inference:} Cryptographic protocols such as MPC and HE secure embeddings and activations traversing WAN links, shielding clinical data during cloud-based or split inference.
Together, these defenses provide partial containment against prompt-level leakage, hidden-state reconstruction, and output correlation, enhancing privacy assurance across deployed healthcare LLMs.
\vspace{-0.1em}\newline
\indent \textcolor{violet}{\textbf{Limitation.}
Despite their value, inference defenses remain fragile when mapped to real-world healthcare deployments:
(1) \textit{Prompt channel and Hidden-State Leakage)}: Local DP blurs sensitive fields but diminishes clinical fidelity; static obfuscation can still be bypassed via cross-session correlation.
(2) \textit{KV cache leakage}:  KV shielding depends on trusted hardware; misconfigured or persistent caches allow recovery of multi-turn patient conversations.
(3) \textit{Returned text (outputs) and Restoration}:  Span-level removal fails under repeated queries, allowing re-identification through prompt–output correlation.
(4) \textit{Delegated / hybrid inference}: MPC and HE secure content but introduce high latency, enabling timing-based inference of prompt complexity or PHI density.
(5) \textit{Operational logging and provenance}: Tracing and debugging pipelines often retain raw prompts and activations, creating durable PHI leakage paths even after session termination.}%

\begin{tcolorbox}[
  colback=gray!5,
  colframe=gray!20,
  boxrule=0.3pt,
  arc=1mm,
  left=2mm,
  right=2mm,
  top=1mm,
  bottom=1mm,
  enhanced,
  breakable,
  before skip=4pt,
  after skip=4pt
]
\textbf{Recommendation 3:}
To address these shortcomings, inference-time privacy must adopt dynamic, context-aware protection tied to each identified surface:\newline
\textbf{(1) Prompt and Hidden-State Protection}: Apply span-level adaptive DP and contextual obfuscation that selectively perturb only sensitive entities (e.g., HIV status, rare conditions), preserving clinical accuracy while blocking re-identification. \textbf{(2) Cross-session correlation Defense}: Introduce session-specific randomization of token mappings and obfuscation schedules to prevent linkage of repeated prompts across oncology or chronic-care cases. \textbf{(3) KV-Cache Security:} Enforce automated cache expiration and TEE-based isolation for multi-turn dialogue memory, preventing GPU scraping or reuse of prior session states. \textbf{(4) Secure Hybrid Inference:}Use hybrid cryptographic inference (MPC + GPU-accelerated HE) to balance latency and confidentiality across WAN-based deployments. \textbf{(5) Operational Logging and Provenance:} Minimize trace capture, anonymize diagnostic logs, and employ privacy-aware provenance tracking to limit long-term PHI persistence. Together, these measures form a phase-aware, layered strategy explicitly mapped to inference-time attack surfaces, closing the residual gaps between theoretical privacy and real-world healthcare deployment safety.
\end{tcolorbox}

Table \ref{tab:inference-taxonomy} presents the inference-phase taxonomy, detailing model-, prompt-, and system-level threats alongside their corresponding defensive strategies, limitations, and recommendations.

\section{Cross-Phase Propagation of Privacy Risks}
Although prior sections analyze vulnerabilities within preprocessing, federated fine-tuning, and inference individually, real-world healthcare deployments rarely operate in isolation. Privacy failures accumulate across phases, allowing seemingly minor leaks in one stage to amplify downstream. Preprocessing artifacts such as residual identifiers, demographic signals, or mislabeled data can be memorized during fine-tuning and later resurfaced through inference-time extraction attacks. Conversely, insecure inference interfaces (e.g., prompt injection, KV-cache probing) can exfiltrate model representations that encode sensitive patterns originating from earlier phases. This lifecycle coupling explains why phase-specific defenses: differential privacy in training, anonymization in preprocessing, or filtering in inference often fail when deployed independently.

Propagation also creates indirect vulnerabilities. Poisoned or synthetic records introduced during preprocessing can bias model gradients during fine-tuning, which then distort inference behavior in clinically consequential ways. Similarly, unsecured logs, caching mechanisms, or checkpoint reuse allow auxiliary metadata from later stages to re-identify samples that were previously anonymized. Threats therefore move bidirectionally: upstream preprocessing influences representational leakage downstream, while downstream inference interfaces expose training-time weaknesses upstream. This interconnectedness underscores a key observation of this SoK: protecting any single phase without accounting for its interactions with the others is structurally insufficient.

This insight motivates the need for lifecycle-aware privacy mechanisms, which the conclusion elaborates by integrating cross-phase findings into a coherent set of deployment-ready recommendations.

\section{Conclusion}
\vspace{-0.3em}
This work presents the first phase-aware systematization of privacy and security in healthcare LLMs, tracing risks and defenses across data preprocessing, fine-tuning, and inference. By mapping adversaries, capabilities, attack surfaces, and defenses at each stage, we illustrate how privacy risks manifest uniquely across the LLM lifecycle and why phase-agnostic approaches fall short in clinical settings

Our analysis highlights a central gap: existing techniques, from anonymization and DP to secure aggregation and cryptographic inference, provide protection only in isolation. In practice, they struggle with \textbf{rare-disease exposure}, \textbf{multi-round reconstruction}, \textbf{poisoning}, and \textbf{runtime leakage} through logs, caches, and distributed infrastructure. Effective privacy in healthcare LLMs therefore requires defenses that are \textbf{context-aware}, \textbf{workflow-aligned}, and \textbf{robust across phases}, not just individually strong.
We identify three priorities for future research: 
(1) \textbf{layered, adaptive defenses} that combine DP, randomized adapters, and hardware enclaves with phase-specific tuning; 
(2) \textbf{standardized evaluation and auditing protocols} for jointly assessing privacy and clinical fidelity; and 
(3) \textbf{lightweight secure computation and federated inference} that meet the latency and reliability demands of real clinical deployments.

By structuring a fragmented landscape into a unified, phase-aware framework, this SoK provides a clear roadmap for developing privacy-resilient, regulation-aligned, clinically safe LLM systems that can earn trust in high-stakes healthcare settings.

\bibliographystyle{IEEEtran}
\bibliography{references}

\end{document}